\documentclass[showpacs,showkeys,11pt,
preprint,preprintnumbers,nofootinbib,
groupedaddress,superscriptaddress,amsmath,amssymb]{revtex4}

\usepackage[dvipdfm]{graphicx}
\usepackage{cancel}
\usepackage[hypertex]{hyperref}

\begin{document}
\title{The Higgs boson mixes with an $SU(2)$ septet representation}
\preprint{IPMU13-0022}
\author{Junji Hisano}
\email{hisano@eken.phys.nagoya-u.ac.jp}
\affiliation{Department of Physics, Nagoya University, Nagoya 464-8602, Japan}
\affiliation{IPMU, TODIAS, University of Tokyo, Kashiwa 277-8568, Japan}
\author{Koji Tsumura}
\email{ko2@eken.phys.nagoya-u.ac.jp}
\affiliation{Department of Physics, Nagoya University, Nagoya 464-8602, Japan}

\date{\today}

\begin{abstract}
We study a possibility of the Higgs boson, which consists of an $SU(2)$ doublet and a septet. 
The vacuum expectation value of a septet with hypercharge $Y=2$ is known to preserve 
the electroweak rho parameter unity at the tree level. 
Therefore, the septet can give significant contribution to the electroweak symmetry breaking. 
Due to the mixing with the septet, the gauge coupling of the standard-model-like Higgs boson 
is larger than that in the standard model. 
We show the sizable VEV of the Higgs septet can be allowed under the constraint from 
the electroweak precision data.
The signal strengths of the Higgs boson for the diphoton and 
a pair of weak gauge boson decay channels at the LHC are enhanced, 
while those for the fermonic decay modes are suppressed. 
The mass of additional neutral Higgs boson is also bounded by the current LHC data 
for the standard model Higgs boson. 
We discussed the phenomenology of the multiply charged Higgs bosons, 
which come from the septet.
\end{abstract}
\maketitle

\section{Introduction}
The Higgs boson has been discovered at the LHC~\cite{Ref:atlas,Ref:cms}. 
However, the detailed information such as interaction strengths with each 
particle have not yet been tested very precisely. 
The most evident signal excess compared with the background 
at the both experiments  
(ATLAS~\cite{Ref:atlas-combi} and CMS~\cite{Ref:cms-combi}) 
has been observed in the diphoton decay channel of the Higgs boson. 
So far the reported rates of this channel are larger than the value predicted 
in the standard model (SM). 
The second best excess was discovered by $ZZ (\to 4\ell)$ decay mode. 
We have also seen the small excess in the $WW$ decay channel. 
The fermonic decay modes of the Higgs boson have not yet established directly. 

The Higgs sector of the SM is still uncertain, which can be extended in various ways. 
One of the promising extension is the supersymmetry, which is motivated 
by the stability of the Higgs boson mass under the quantum corrections. 
In this model, the Higgs sector is required to have even number of the doublet. 
Another interesting extension is the Higgs triplet model~\cite{Ref:Type-II}, 
which consists of the SM Higgs doublet and a Higgs triplet with $Y=1$. 
In this model, Majorana neutrino masses are generated from the small 
vacuum expectation value (VEV) of the triplet. 
Since there are many extensions of the SM, we need to discriminate 
them from each other at further precision measurement of the Higgs boson. 

The Veltman's rho parameter~\cite{Ref:rho} has been measured very precisely.
It would be a good guiding principle to construct the beyond the SM. 
For an arbitrary number of $SU(2)$ scalar multiplets, 
which have an isospin $j_\phi$, a hypercharge $Y_\phi$ and a VEV $v_\phi$, 
the rho parameter at tree level is calculated as
\begin{align}
\rho &=
\frac{\Sigma_\phi\, [\, j_\phi^{}(j_\phi^{}+1)-Y_\phi^2\, ]\, v_\phi^2}{
\Sigma_\phi\, 2Y_\phi^2\, v_\phi^2}.
\end{align}
The precision measurement suggests $\rho$ is very close to unity. 
In the SM, the Higgs doublet ($j_\phi^{}=\frac12$ and $Y_\phi=\tfrac12$) predicts $\rho=1$. 
The multi doublet extensions of the SM, e.g., the minimal supersymmetric standard model, 
also predict $\rho=1$  at tree level. 
It is known that the next minimal representation, which satisfy $\rho=1$, is  a septet 
with $Y_\phi = 2$, and the next next minimal one is a 26-plet 
with $Y_\phi= \frac{15}2$. 
%
In the usual Higgs representation, the VEV is bounded to be small. 
For instance, in the Higgs triplet model which consists of a doublet and a triplet Higgs fields, 
the VEV of the triplet is constrained to be less than 3 GeV ($2\sigma$ level) 
from $\rho_0 = 1.0004^{+0.0003}_{-0.0004}$~\cite{Ref:pdg}, 
where $\rho_0 = \rho/\rho_\text{SM}$.
If we allow the cancellation among various contributions, 
the rho parameter can remain unity~\cite{Ref:GM}.

In this paper, we construct a framework of the low energy effective theory, 
which contains a Higgs doublet and a Higgs septet with sizable VEVs.  
The electroweak sector of the model is extended due to the presence of 
the septet, while keeping the rho parameter unity at tree level. 
The model is consistent with the electroweak precision data based on 
the $S$ and $T$ parameter analysis. 
The gauge coupling of the SM-like Higgs boson can be largely deviated 
from the SM value due to the mixing with the septet component. 
The signal strengths of the Higgs boson for the diphoton and 
for the weak gauge boson pair decays can be enhanced, 
while those for fermonic decay channels are suppressed. 
Additional Higgs bosons appear in the weak scale and would be discovered at the LHC. 
In particular, the singly charged Higgs boson can be produced by the $W^\pm Z$ fusion mechanism. 

This paper is organized as follows. 
In Section II, we present a model which consists of a Higgs doublet and a Higgs septet. 
The electroweak precision constraints, and the signal strengths of the Higgs bosons 
at the LHC are studied in Section III. 
The characteristic signatures of the extra (septet-like) Higgs boson are discussed. 
Conclusions and discussion are given in Section IV. 

\section{The model}
A scalar septet field $(\chi)$ with $Y=2$ is introduced to the SM 
in addition to the SM Higgs doublet $(\Phi)$ with $Y=1/2$.
The Higgs potential is given by 
\begin{align}
{\mathcal V}
=& -\mu^2_2 \Phi^\dag\Phi +M_7^2 \chi^\dag\chi
+\lambda (\Phi^\dag\Phi)^2
-\frac1{\Lambda^3} \{ (\chi \Phi^5 \Phi^*) + \text{H.c.} \}
\nonumber \\
&\quad
+\sum_{A=1}^4 \lambda_A (\chi^\dag \chi \chi^\dag \chi)_A 
+\sum_{B=1}^2 \kappa_B (\Phi^\dag \Phi \chi^\dag \chi)_B
, 
\end{align}
where
\begin{align}
(\Phi^\dag \Phi \chi^\dag \chi)_1 
&= \phi^{*i} \phi^{}_i \chi^{*abcdef} \chi^{}_{abcdef},\\
(\Phi^\dag \Phi \chi^\dag \chi)_2
&= \phi^{*i} \phi^{}_j \chi^{*jabcde} \chi_{iabcde},\\
(\chi^\dag \chi \chi^\dag \chi)_1 
&= \chi^{*ijklmn} \chi^{}_{ijklmn} \chi^{*abcdef} \chi^{}_{abcdef},\\
(\chi^\dag \chi \chi^\dag \chi)_2
&= \chi^{*ijklmn} \chi^{}_{ijklmf} \chi^{*abcdef} \chi^{}_{abcden},\\
(\chi^\dag \chi \chi^\dag \chi)_3 
&= \chi^{*ijklmn} \chi^{}_{ijklef} \chi^{*abcdef} \chi^{}_{abcdmn},\\
(\chi^\dag \chi \chi^\dag \chi)_4 
&= \chi^{*ijklmn} \chi^{}_{ijkdef} \chi^{*abcdef} \chi^{}_{abclmn},
\end{align}
and 
\begin{align}
(\chi^* \Phi^5 \Phi^*) = \chi^{abcdef} \Phi_a \Phi_b \Phi_c \Phi_d \Phi_e \Phi^{*g} \epsilon_{fg}
\end{align}
with $\Phi_1=\omega_2^+$ and $\Phi_2=(v_2+h_2+i\,z_2)/\sqrt2$. 
We here introduce the symmetric tensor notation of the septet field as
$\chi_{111111}=\chi_3$, 
$\chi_{111112}=\chi_2/\sqrt{6}$,
$\chi_{111122}=\chi_1/\sqrt{15}$,
$\chi_{111222}=\chi_0/\sqrt{20}$,
$\chi_{112222}=\chi_{-1}/\sqrt{15}$,
$\chi_{122222}=\chi_{-2}/\sqrt{6}$, and 
$\chi_{222222}=\chi_{-3}$
with 
$\chi_{-2}=(v_7+h_7+i\,z_7)/\sqrt2$. 
Although the Higgs septet can break the electroweak symmetry properly, 
the Higgs doublet is required for the mass generation of SM fermions.
Without a non-renormalizable term with mass dimension seven, 
$\chi^* \Phi^5 \Phi^*$, the Higgs potential is conserved under 
an accidental global $U(1)$ transformations of the doublet and the septet, 
separately. 
Therefore, a breaking term of the global symmetry is added 
in order to avoid an exact massless Nambu-Goldstone boson.  

\begin{figure}[tb]
 \centering
 \includegraphics[height=5.cm]{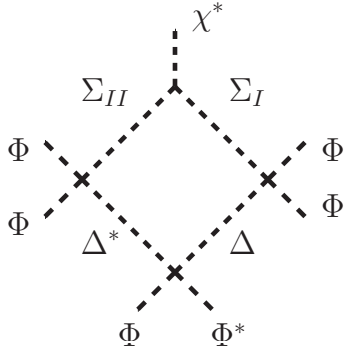} 
 \caption{Radiative generation of dimension seven operator, $\chi^* \Phi^5 \Phi^*$,  
 from the renormalizable UV theory.}
 \label{FIG:dim7}
\end{figure}

As an example, the dimension seven operator can be generated 
from the one loop contribution given in FIG.~\ref{FIG:dim7} 
by introducing two scalar quintuplets $\Sigma_{I, II} (Y=1)$ 
and a scalar triplet $\Delta (Y=0)$,  which are odd under 
the extra discrete symmetry. 
The renormalizable Lagrangian related to this one loop diagram is given by
\begin{align}
{\mathcal L}_{\cancel{U(1)}}
= \mu\,  \chi_{abcdef} \Sigma_I^{*abci} \Sigma_{II}^{*defj} \epsilon_{ij}
+\Phi_i \Phi_j (c_I^{}\, \Sigma_I^{*ijkl} +c_{II}^{}\, \Sigma_{II}^{*ijkl})\Delta_{kl}  
+f\, \Phi_a \Phi^{*b} \Delta^{*ac} \Delta_{bc} +\text{H.c.},
\end{align}
 where $\mu$ is a soft breaking mass parameter of the global $U(1)$ symmetry, 
 and $c_{I, II}^{}$ and $f$ are the coupling constants. 
 In the following discussion, we take into account the dimension seven operator 
 instead of specifying the UV completion of the model 
 without limiting the generality of the low energy effective theory\footnote{
 Phenomenology of the UV theories including the dark matter will be studied~\cite{Ref:HT}.
 }.  

If $M_7^2 < 0$, the Higgs septet and the Higgs doublet develop the VEVs independently. 
In this case, there are possibilities of the charge breaking vacua. In addition, 
the stability of the vacuum is unclear even in the charge conserving vacuum.  In this paper, we consider the case with $M_7^2 >0$, where the VEV of the Higgs septet 
is induced through the higher dimensional operator,
\begin{align}
&v_7 \simeq \tfrac{\sqrt3}{24}\, \tfrac{v_2^6}{\Lambda^3 M_7^2}. 
\label{Eq:VC}
\end{align}
The VEVs of the Higgs doublet and septet are automatically aligned so that 
the charge conserving vacuum is realized. Further, we assume that $\lambda_{1-4}$ and 
$\kappa_{1,2}$ in the Higgs potential are sufficiently small for simplicity.
In this case, we can parametrize the model by only two parameters with 
the SM-like Higgs boson mass as shown later.
Let us remind you that the septet quartic coupling constants generically 
produce mass splittings among the septet-like scalar bosons, and the latter 
coupling constants generate mixings between the doublet and the septet. 
In the following discussion, we mostly neglect the effect of $\lambda_{1-4}$ 
and $\kappa_{1,2}$, but a few important implications will be mentioned 
in the relevant parts.

The mass matrices for the scalar sector are diagonalized by 
\begin{align}
\begin{pmatrix} h_7 \\ h_2\end{pmatrix} 
&= 
\begin{pmatrix} c_\alpha  & -s_\alpha\\
s_\alpha & c_\alpha \end{pmatrix}
\begin{pmatrix} H \\ h\end{pmatrix}, \\
\begin{pmatrix} z_7 \\ z_2\end{pmatrix} 
&= 
\begin{pmatrix} c_\beta  & -s_\beta\\
s_\beta & c_\beta \end{pmatrix}
\begin{pmatrix} z \\ A\end{pmatrix}, \\
\begin{pmatrix} \chi_{-1} \\ \chi_{-3}^* \\ \omega_2^+ \end{pmatrix} 
&= 
\begin{pmatrix} 
\tfrac{\sqrt{10}}4 & \tfrac{\sqrt6}4 & 0 \\
-\tfrac{\sqrt6}4 & \tfrac{\sqrt{10}}4 & 0 \\ 
0 & 0 & 1 \end{pmatrix}
\begin{pmatrix} c_\beta & 0 & -s_\beta \\
0 & 1 & 0 \\ 
s_\beta & 0 & c_\beta \end{pmatrix}
\begin{pmatrix} \omega^+ \\ H_2^+ \\ H_1^+\end{pmatrix},
\end{align}
where $z$ and $\omega^\pm$ are the electroweak Nambu-Goldstone bosons, 
the ratio of VEVs, $\tan\beta = v_2/(4 v_7)$, and 
the CP even scalar boson mixing, $\alpha$. 
Note that the doublet VEV is dominated in the large $\tan\beta$ limit. 
Mass eigenvalues are calculated as 
\begin{align}
m_h^2
=& (1+ \tfrac32\, \tfrac1{t_\beta\, t_\alpha})\, M_7^2, \\
m_H^2
=& (1- \tfrac32\, \tfrac{t_\alpha}{t_\beta})\, M_7^2, \\
m_A^2 
=& m_{H_1^\pm}^2 = M_7^2/s_\beta^2, \\
m_{H_2^\pm}^2 
=& m_{H^{2\pm}}^2 = m_{H^{3\pm}}^2 = m_{H^{4\pm}}^2 = m_{H^{5\pm}}^2 
= M_7^2,
\end{align}
where $\chi_3=H^{5+}$, $\chi_2=H^{4+}$, $\chi_1=H^{3+}$, $\chi_0=H^{2+}$.
There are four input parameters.  
We here choose $v=(\sqrt2 G_F)^{-1/2}$, $m_h = 126$ GeV, $M_7$ and $\tan\beta$. 
The electroweak interactions of these scalar bosons are determined straightforwardly. 

\begin{figure}[tb]
 \centering
 \includegraphics[height=6.cm]{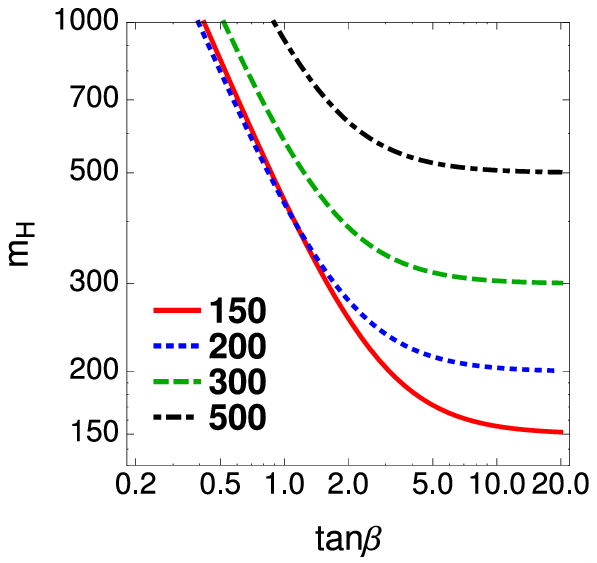} 
 \includegraphics[height=6.cm]{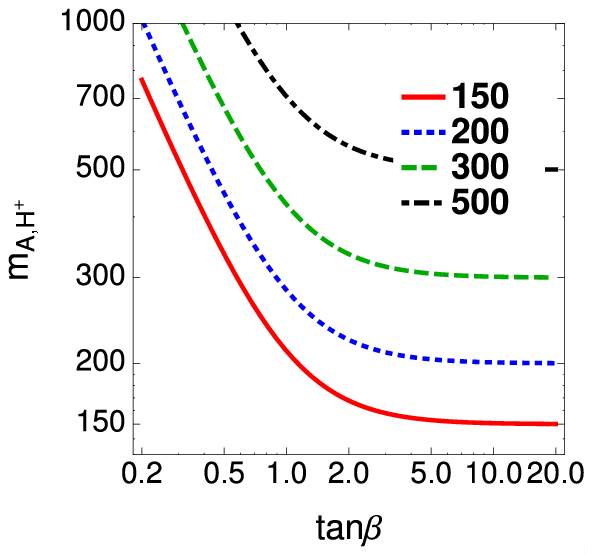} 
 \caption{Masses for $H$ (left), $A$ and $H_1^\pm$ (right) are given 
 as functions of $\tan\beta$. 
 The solid, dashed, dotted and dot-dashed curves correspond to 
 $M_7=150, 200, 300$ and $500$ GeV, respectively.}
 \label{FIG:Mass}
\end{figure}

In FIG.~\ref{FIG:Mass}, masses for $H$ (left), $A$ and $H_1^\pm$ (right)
are shown as functions of $\tan\beta$. 
The solid, dashed, dotted and dot-dashed curves denote 
the cases for $M_7=150, 200, 300$ and $500$ GeV, respectively. 
They coincide with $M_7$ value at the large $\tan\beta$ region. 
Mass splittings can be large in the small $\tan\beta$ region, 
and $m_H^{} > m_A^{}$ is satisfied. 
We can see from Eq.~\eqref{Eq:VC} the UV completion of 
the dimension seven operator spoils rapidly for $\tan\beta \ll 1$. 
As for the reference, the values of $m_H^{}$ and $m_A^{}$ are listed in TABLE.~\ref{Tab:Mass}.

\begin{table}[tb]
  \begin{tabular}{c||cccc}
  $(m_H^{}, m_A^{})$ [GeV]
    & $\tan\beta=3$ & $\tan\beta=5$ & $\tan\beta=10$ & $\tan\beta=20$ \\
   \hline \hline
    $M_7=150$ GeV 
    & (204., 158.) & (171., 153.) & (156., 151.) & (151., 150.) \\
    $M_7=200$ GeV
    & (238., 211.) & (214., 204.) & (204., 201.) & (202., 200.)\\
    $M_7=300$ GeV
    & (343., 316.) & (316., 306.) & (304., 301.) & (302., 301.)\\
    $M_7=500$ GeV
    & (563., 527.) & (523., 510.) & (506., 502.) & (503., 501.)
  \end{tabular}
  \caption{Masses of the heavy CP even and CP odd Higgs bosons  
  are listed for $M_7 = 150, 200, 300$ and $500$ GeV.}
  \label{Tab:Mass}
\end{table}

\begin{figure}[tb]
 \centering
 \includegraphics[height=6.5cm]{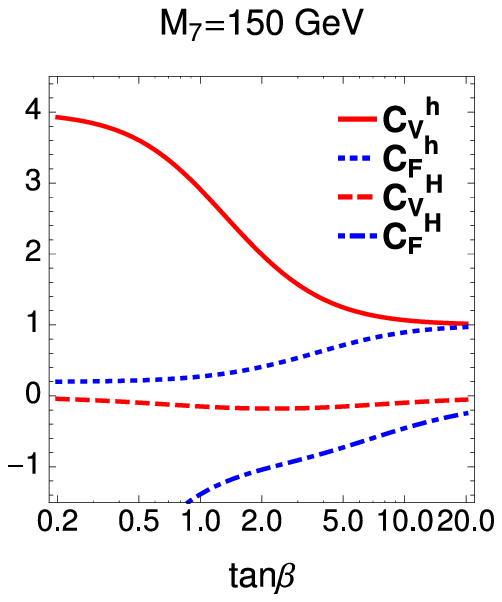} 
 \includegraphics[height=6.5cm]{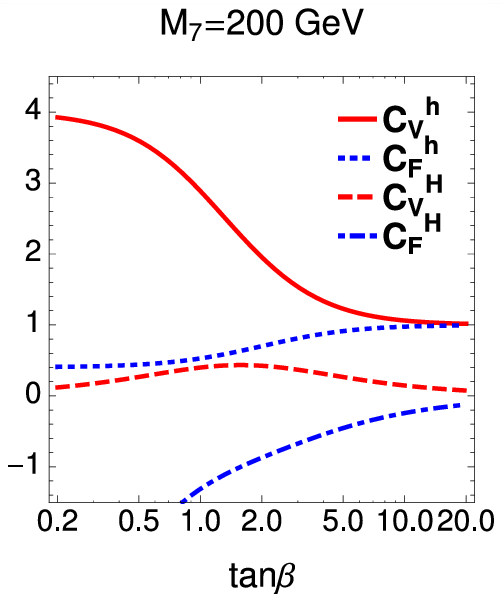} 
 \includegraphics[height=6.5cm]{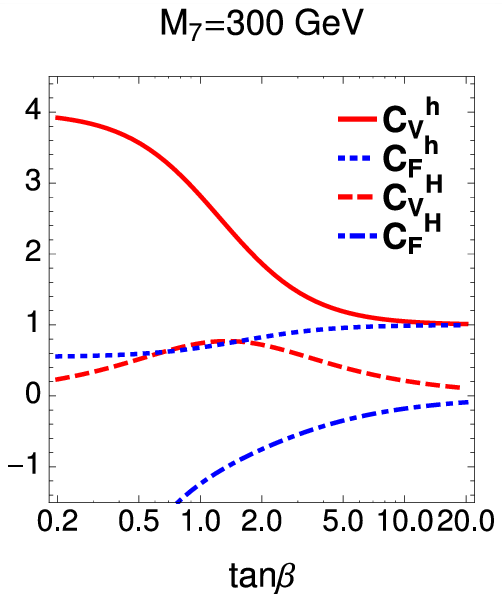} 
 \includegraphics[height=6.5cm]{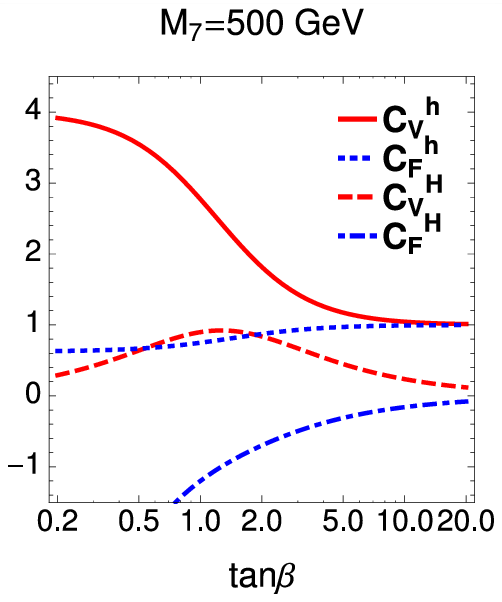} 
 \caption{Correction factors for the gauge and the Yukawa coupling constants 
 of the CP even Higgs bosons are presented  as functions of $\tan\beta$ 
 for $M_7=150$ GeV (top-left), $M_7=200$ GeV (top-right), 
 $M_7=300$ GeV (bottom-left) and $M_7=500$ GeV (bottom-right). 
 The solid, dotted, dashed and dot-dashed curves express $C_V^h$, $C_F^h$, 
 $C_V^H$ and $C_F^H$, respectively. 
 }
 \label{FIG:CVCF}
\end{figure}

Let us discuss the modification of the Higgs coupling constant. 
The correction factors for the gauge interaction of the CP even Higgs bosons are given by
\begin{align}
C_V^h &= s_\beta\, c_\alpha - 4\, c_\beta\, s_\alpha, \\
C_V^H &= s_\beta\, s_\alpha + 4\, c_\beta\, c_\alpha.
\end{align}
The factors are normalized by the SM Higgs boson coupling constant, 
i.e., $C_V^\phi \equiv \lambda_{\phi VV}/\lambda_{H_\text{SM}^{}VV}$. 
Similar correction factors for the Yukawa interaction are also given by
\begin{align}
C_F^h &= c_\alpha/s_\beta, \\
C_F^H &= s_\alpha/s_\beta.
\end{align}
These expressions are the same as in the so-called Type-I two Higgs doublet model~\cite{Ref:HHG}.

In FIG.~\ref{FIG:CVCF}, we show the modification factors, $C_V^\phi$ and $C_F^\phi$, 
of the CP even Higgs boson $(\phi=h, H)$ for $M_7=150$ GeV (top-left), 
$M_7=200$ GeV (top-right), $M_7=300$ GeV (bottom-left) and $M_7=500$ GeV (bottom-right). 
The solid, dotted, dashed and dot-dashed curves represent $C_V^h$, $C_F^h$, 
 $C_V^H$ and $C_F^H$, respectively. 
 Since the doublet VEV is dominated in the large $\tan\beta$ region, the coupling strengths 
of $h$ become SM-like, while those of $H$ vanish. 
The maximal value of $C_V^h$ is four, which is determined by the hypercharges 
of the septet and the doublet.
The gauge interaction of the SM-like Higgs boson is greater than one for the all plots 
because of the mixing with the septet component. 

\section{Phenomenology}

In this section, we investigate the current experimental constraint 
from the electroweak precision data and also from the Higgs boson 
searches at the LHC. 
The LHC signatures of the model are also studied.

We begin with the constraint from the electroweak sector. 
In this model, the oblique corrections~\cite{Ref:PeskinTakeuchi} from 
the scalar bosons are calculated as
\begin{align}
S =& \frac1{4\pi} \bigl[
(C_V^h)^2\, {G^{hZ}}'
+(C_V^H)^2\, {G^{HZ}}'
+30\, c_\beta^2\, {G^{H_2^\pm W}}'
+30\, s_\beta^2\, {F^{H_1^\pm H_2^\pm}}'
 \nonumber \\
 &\qquad
 -\tfrac13\, \ln m_{H_1^\pm}^2 -15\, \ln m_{H_2^\pm}^2
 +(4\, s_\alpha s_\beta -c_\alpha c_\beta) {F^{hA}}'
 +(4\, c_\alpha s_\beta +s_\alpha c_\beta) {F^{HA}}'
\bigr],\\
T =& 
\frac{\sqrt2G_F}{\alpha_\text{EM}^{}(4\pi)^2} \bigl[ 
(C_V^h)^2\, \Delta G^{h}
+(C_V^H)^2\, \Delta G^{H}
-15\, c_\beta^2\, \Delta G^{H_2^\pm}
\bigr], 
\end{align}
where the loop functions, ${G^{xy}}'$, ${F^{xy}}'$, and $\Delta G^{x}$, are given in Appendix.

\begin{figure}[tb]
 \centering
 \includegraphics[height=6.3cm]{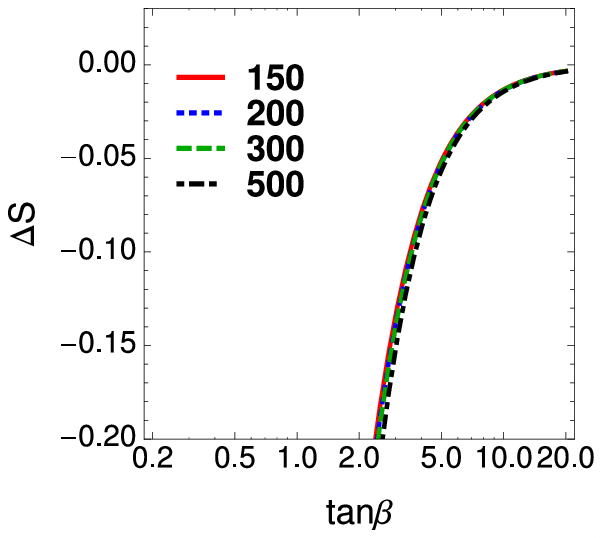} 
 \includegraphics[height=6.cm]{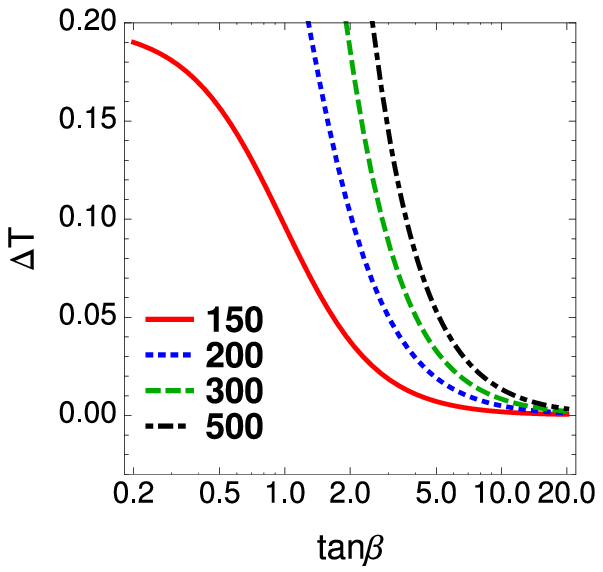} 
 \caption{Oblique corrections are evaluated as functions of $\tan\beta$.
 The solid, dotted, dashed curves represent the cases 
 for $M_7=150$, $200$ and $300$ GeV, respectively.
 }
 \label{FIG:EWPO}
\end{figure}

\begin{table}[tb]
  \begin{tabular}{c||cccc}
  $(\Delta S, \Delta T)$
    & $\tan\beta=3$ & $\tan\beta=5$ & $\tan\beta=10$ & $\tan\beta=20$ \\
   \hline \hline
    $M_7=150$ GeV 
    & (-0.13, 0.019) & (-0.05, 0.007) & (-0.013, 0.002) & (-0.003, 0.) \\
    $M_7=200$ GeV
    & (-0.14, 0.050) & (-0.05, 0.019) & (-0.014, 0.005) & (-0.003, 0.001)\\
    $M_7=300$ GeV
    & (-0.14, 0.088) & (-0.05, 0.033) & (-0.013, 0.008) & (-0.003, 0.002)\\
    $M_7=500$ GeV
    & (-0.15, 0.14) & (-0.06, 0.053) & (-0.014, 0.013) & (-0.004, 0.003)
  \end{tabular}
  \caption{Oblique corrections to $S$ and $T$ parameters  
  are listed for $M_7 = 150, 200, 300$ and $500$ GeV.}
  \label{Tab:EWPO}
\end{table}

In FIG.~\ref{FIG:EWPO}, the deviations of the $S$ parameter (left) and 
$T$ parameter (right) are plotted 
for $M_7=150$ GeV (solid), $M_7=200$ GeV (dotted), $M_7=300$ GeV (dashed) 
and $M_7=500$ GeV (dot-dashed) as functions of $\tan\beta$. 
The current experimental bounds on oblique parameters are $\Delta S=0.04 \pm 0.09$ 
and $\Delta T=0.07 \pm 0.08$ with the $88\%$ correlation by fixing $\Delta U=0$~\cite{Ref:pdg}. 
In the small $\tan\beta$ region, 
relatively larger negative (positive) contributions to the $S$ $(T)$ parameter are found. 
For $\Delta T=0$, $\Delta S$ is constrained to be larger than $-0.08\, (-0.11)$ 
at $68\, (90) \%$ confidence level. 
From the $S$ parameter only, $\tan\beta \gtrsim 5$ is obtained. 
Because we are fixing the value of $\tan\beta$, these effects are non-decoupling for higher $M_7$. 
In fact, the largest contribution to the $T$ parameter is seen for $M_7=500$ GeV. 
As for the reference, the values of $\Delta S$ and $\Delta T$ are listed in TABLE.~\ref{Tab:EWPO}.

\begin{figure}[tb]
 \centering
 \includegraphics[height=6.5cm]{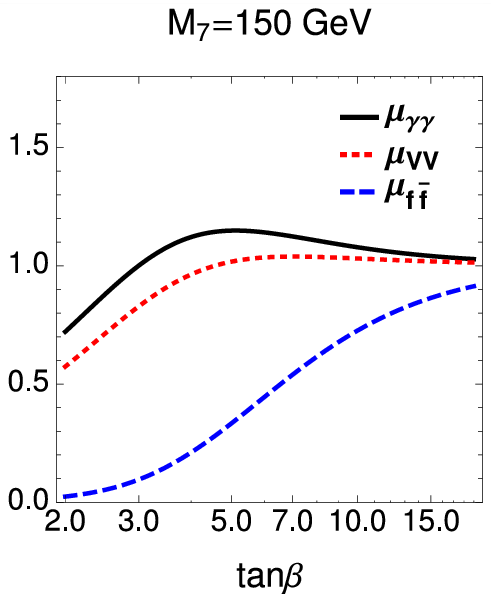} 
 \includegraphics[height=6.5cm]{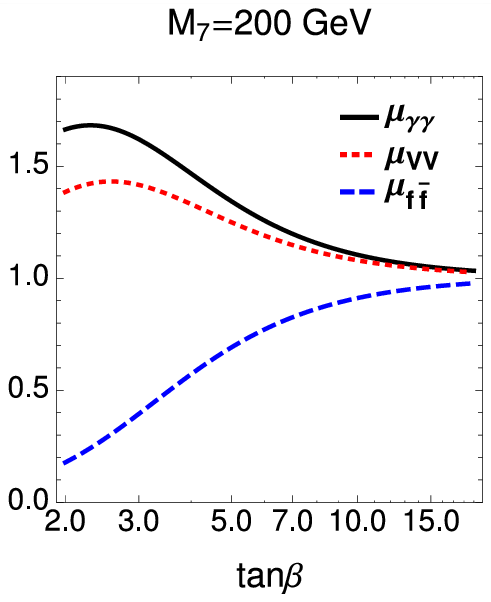} 
 \includegraphics[height=6.5cm]{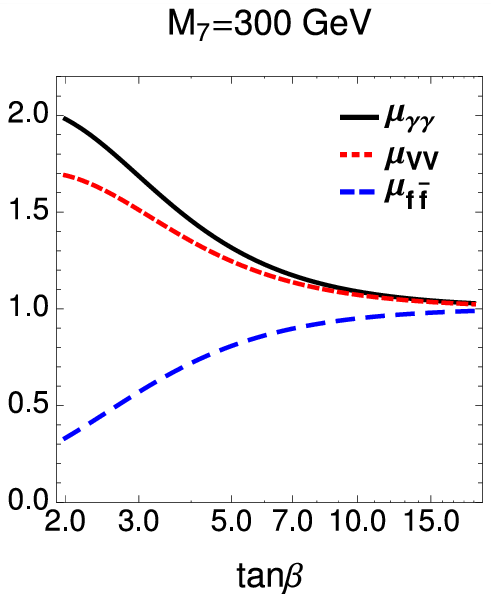} 
 \includegraphics[height=6.5cm]{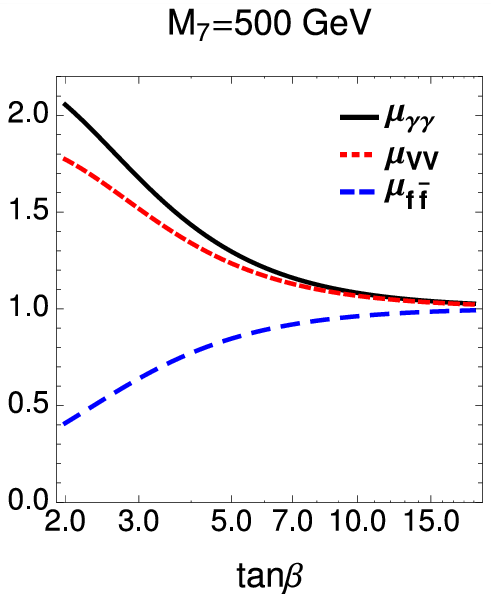} 
 \caption{Signal strengths for each Higgs boson decay mode at the LHC 
 are shown as functions of $\tan\beta$.  
 The mass scale, $M_7$, is chosen as $150$ GeV (top-left), 
 $200$ GeV (top-right), $300$ GeV (bottom-left) and $=500$ GeV (bottom-right).}
 \label{FIG:mu}
\end{figure}

Let us turn to the LHC phenomenology. 
At the LHC, we have found the Higgs boson around 126 GeV using $\gamma\gamma$ 
and $ZZ(\to 4\ell)$ decay channels. 
Since the coupling constants of the SM-like Higgs boson are corrected, 
these signal strengths can be changed. 
Roughly speaking, 90\,\% of the Higgs boson come from the gluon fusion mechanism 
at the LHC with $\sqrt{s} = 7$ TeV and/or $\sqrt{s} = 8$ TeV without selection cuts. 
Assuming the dominance of the gluon fusion production cross section, the signal strengths 
are evaluated as
\begin{align}
\mu_{VV}^h
&\simeq 
(C_F^h)^2 \frac1{
(C_F^{h})^2({\mathcal B}_{bb}+{\mathcal B}_{cc}+{\mathcal B}_{\tau\tau}
+{\mathcal B}_{gg}) +(C_V^{h})^2({\mathcal B}_{WW}+{\mathcal B}_{ZZ}^{})}
(C_V^h)^2, \\
\mu_{f\bar f}^h
&\simeq 
(C_F^h)^2 \frac1{
(C_F^{h})^2({\mathcal B}_{bb}+{\mathcal B}_{cc}+{\mathcal B}_{\tau\tau}
+{\mathcal B}_{gg}) +(C_V^{h})^2({\mathcal B}_{WW}+{\mathcal B}_{ZZ}^{})}
(C_F^h)^2, \\
\mu_{\gamma\gamma}^h
&\simeq
(C_F^h)^2 \frac1{
(C_F^{h})^2({\mathcal B}_{bb}+{\mathcal B}_{cc}+{\mathcal B}_{\tau\tau}
+{\mathcal B}_{gg}) +(C_V^{h})^2({\mathcal B}_{WW}+{\mathcal B}_{ZZ}^{})}
R, 
\end{align}
where 
\begin{align}
R
&=
\Biggl|\frac{C_F^h N_cQ_t^2A_{1/2}(\tau_t^{})+C_V^h A_1(\tau_W^{})}{
N_cQ_t^2A_{1/2}(\tau_t^{})+A_1(\tau_W^{})}\Biggr|^2, \\
A_{1/2}(\tau) 
&= 2 [\tau+(\tau-1)f(\tau)] \tau^{-2}, \\
A_1(\tau)
&= -[2\tau^2+3\tau+3(2\tau-1)f(\tau)] \tau^{-2},
\end{align}
with $\tau_x^{}=m_h^2/(4\, m_x^2)$ and
\begin{align}
f(\tau)=
\begin{cases}
\arcsin^2\sqrt{\tau} & \tau \le 1 \\
-\tfrac14 \Bigl[ \ln\frac{1+\sqrt{1-\tau_{}^{-1}}}{1-\sqrt{1-\tau_{}^{-1}}} -i\, \pi\Bigr]^2 & \tau > 1
\end{cases}.
\end{align}
The recommended values of the branching ratios of the SM Higgs boson are listed 
in TABLE.~\ref{Tab:BR126}~\cite{Ref:XSWG}. 
Because we are neglecting the interactions between the doublet and the septet 
in the Higgs potential, the (multiply) charged Higgs boson loop contributions are 
omitted\footnote{
Detailed studies of the extended scalar sector are beyond the scope of this paper. 
It will be shown in elsewhere~\cite{Ref:HT}. 
}. 

In FIG.~\ref{FIG:mu}, the signal strengths of the SM-like Higgs boson at the LHC are shown 
as functions of $\tan\beta$ for $M_7=150$ GeV (top-left), $M_7=200$ GeV (top-right)
$M_7=300$ GeV (bottom-left) and $M_7=500$ GeV (bottom-right). 
The solid, dotted and dashed curves denote the signal strengths for 
$\gamma\gamma$, $VV^* (V=W,Z)$ and $f\bar{f}$ decays, respectively, of the SM-like Higgs boson, $h$. 
\begin{table}[tb]
  \begin{tabular}{c||cccccc}
    & ${\mathcal B}_{bb}$ & ${\mathcal B}_{cc}$ & ${\mathcal B}_{\tau\tau}$ 
    & ${\mathcal B}_{WW}^{}$ & ${\mathcal B}_{ZZ}^{}$ & ${\mathcal B}_{gg}$ \\
   \hline \hline
    $m_{H_\text{SM}^{}} = 126$ GeV
    & 0.56 & 0.028 & 0.062 & 0.23 & 0.029 & 0.085 \\
  \end{tabular}
  \caption{Recommended value for the Higgs boson decay branching ratios
  for $m_{H_\text{SM}^{}} = 126$ GeV.}
  \label{Tab:BR126}
\end{table}
As we have seen in FIG.~\ref{FIG:CVCF}, the production cross section of 
the Higgs boson via the gluon fusion is suppressed by $(C_F^h)^2$. 
On the other hand, the $b\bar b$ decay channel, which is the main decay mode 
of the SM Higgs boson, is suppressed in the present model. 
Furthermore, the effective gauge coupling of the SM-like Higgs boson is 
larger than that of the SM Higgs boson. 
Therefore, the net contributions to $\mu_{VV}^h$ can be enhanced.  
The diphoton decay mode is more enhanced than $VV$ mode due to 
the enhanced gauge interaction and the suppressed Yukawa interaction in the loop. 
For $M_7=150$ GeV, $\mu_{\gamma\gamma}^h$ can be close to 1.2 with 
$\tan\beta \simeq 5$, which is allowed by electroweak precision data. 
In this case, $\mu_{VV}^h$ is close to unity, which is favored 
by the current experimental data. 
For larger $M_7$, $\mu_{\gamma\gamma}^h$ can be larger than 1.5 
in the small $\tan\beta$ region, 
which is however excluded by electroweak precision data. 

It is too early to discuss the precise values of the signal strengths. 
Qualitatively, the larger signal strengths for the bosonic decay modes, and 
the smaller signal strengths for the fermonic decay modes are predicted. 
You might consider that the enhancement of the diphoton decay rate is not sufficient 
to explain the observations at the LHC. 
However, in the Higgs septet there are many multiply charged Higgs bosons, 
which may further enhance diphoton signal if there are interactions between 
the doublet and the septet. 

The current LHC bound for the heavier SM Higgs boson can be rescaled 
to constrain the mass of $H$ using $C_F^H$ and $C_V^H$ factors. 
For the heavier CP even Higgs boson, the strongest bound comes from 
the signal strength of weak gauge boson decay modes. 
Using the recommended values of the SM Higgs decay branching ratios, 
the signal strength is evaluated in TABLE.~\ref{Tab:mu}, 
where only the decay modes into the SM particles are included. 
The decays of $H\to AZ^*$ and $H\to H_1^\pm W^{\mp*}$ give 
negligible contribution due to the small mass splittings. 
From the LHC data~\cite{Ref:atlas-Dec4l,Ref:cms-combi}, 
the signal strength is constrained to be less than 0.1--0.3 
for $m_H^{} = 150$--$500$ GeV.
For $M_7 \simeq 200$ GeV, $\tan\beta \gtrsim 5$ is allowed by 
current experimental data. 
For smaller $M_7$, $\tan\beta$ is required to be small 
in order to avoid the constraint from the non-discovery of such a boson. 

\begin{table}[tb]
  \begin{tabular}{c||c|c|c|c|c|c}
  ($m_H^{}$[GeV], $\mu_{VV}^H$)
    & $\tan\beta=5$ & $\tan\beta=6$ & $\tan\beta=7$ & $\tan\beta=8$ & $\tan\beta=9$ & $\tan\beta=10$  \\
   \hline \hline
    $M_7=150$ GeV & (171., 0.44) & (165., 0.31) & (161., 0.20) 
                               & (159., 0.13) & (157., 0.081) & (156., 0.062) \\
    $M_7=200$ GeV & (214., 0.21) & (210., 0.15) & (207., 0.11) 
                               & (206., 0.089) & (205., 0.071) & (204., 0.059) \\
    $M_7=300$ GeV & (316., 0.12) & (311., 0.087) & (308., 0.065) 
                               & (306., 0.050) & (305., 0.040) & (304., 0.032) \\
    $M_7=500$ GeV & (523., 0.12) & (516., 0.084) & (512., 0.063) 
                               & (509., 0.048) & (507., 0.038) & (503., 0.031)
  \end{tabular}
  \caption{Signal strength for $H \to VV$ decay modes
   are listed for $M_7 = 150, 200, 300$ and $500$ GeV.}
  \label{Tab:mu}
\end{table}

The smoking gun of the septet contribution would be the presence of 
the $Z W^\mp H^\pm$ vertex. 
This kind of the interaction is prohibited in the multi Higgs doublet models, 
which give $\rho=1$ at tree level. 
The higher dimensional representation can generate this vertex, 
however the size of the coupling strength is suppressed by the small VEV 
due to the strong constraint from the rho parameter. 
Since the VEV of the Higgs septet with $Y=2$ can be sizable, 
relatively large $Z W^\mp H^\pm$ vertex is allowed. 
In this model, we have 
\begin{align}
{\mathcal L}_v =
&
+ \tfrac{\sqrt{15}}2 \, g\, g_Z^{}\, v\, c_\beta Z_\mu \bigl( 
H_2^- W^{+\mu} +H_2^+  W^{-\mu} \bigr).
\end{align}
Note that if the term proportional to $\kappa_i$ exists in the Higgs potential, 
$H_1^\pm$ also couples with $Z W^\mp$ via the mixing with $H_2^\pm$.  
If the charged Higgs boson $H_2^\pm$ is sufficiently light and $\tan\beta \lesssim 10$, 
$H_2^\pm$ can be produced by $W^\pm Z$ fusion at the LHC~\cite{Ref:WZH-lhc}. 
The precision measurement of this coupling is possible at the ILC 
using the recoil method via $e^+e^- \to Z^* \to W^\mp H_2^\pm$~\cite{Ref:WZH-ilc}.

The presence of the multiply charged Higgs bosons is also an interesting feature of the model. 
Due to the relatively large VEV of the septet, the doubly charged Higgs boson in this model 
mostly decays into the same-signed weak gauge bosons and is produced via 
$W^\pm W^\pm \to H^{2\pm}$ process. 
Such a process has been studied~\cite{Ref:CNT} in the context of the Higgs triplet model 
and the Georgi-Machacek model.
The multiply charged Higgs boson with electric charge higher than two has 
a long decay chain, which provide multiple weak boson final states.  

\section{Conclusions}

We have studied the model consists of the Higgs doublet with $Y=1/2$ and 
the Higgs septet with $Y=2$. 
The VEVs of the doublet and the septet preserve the electroweak rho parameter unity 
at tree level without any cancellation. 
The presence of the Higgs septet can give substantial contributions to 
the electroweak symmetry breaking. 
A non-renormalizable operator is introduced in order to break the accidental global 
symmetry. A UV completion of the operator is also presented. 
From the view point of the UV completion, 
it would be difficult to introduce much higher dimensional representations with 
a sizable VEV such as 26-plet with $Y=\frac{15}2$. 
Therefore, the Higgs septet would only be a realistic possibility besides doublets

In this model, the gauge coupling constant of the SM-like Higgs boson $(h)$ 
is predicted to be larger than that in the SM.  
We have shown the modified electroweak interactions due to the sizable VEV 
of the Higgs septet are consistent with the electroweak precision data. 
The signal strengths of the SM-like Higgs boson at the LHC
for the diphoton and for the weak boson pair decay modes 
are enhanced, while those for the fermonic decay channels are suppressed. 
The mass of the additional CP even neutral Higgs boson $(H)$ is constrained 
by the current LHC data depending on the ratio of the VEVs. 
The VEV of the Higgs septet induces sizable $W^\mp Z H^\pm$ vertex ($
H^\pm$ is the charged Higgs boson in the Higgs septet), 
which is forbidden in the multi Higgs doublet extension of the SM and 
is highly suppressed by the VEV. e.g., in the Higgs triplet model. 
One of the interesting signature of the model would be the $W^\pm Z$ 
fusion production of the charged Higgs boson at the LHC. 

The electroweak sector of the model has been discussed in this paper. 
We would like to mention the implications of the model. 
As we have slightly noted the diphoton decay rate of the SM-like Higgs boson 
can be enhanced by the effects of the multiply charged Higgs boson loops. 
Such a effect generally requires larger coupling constants in the Higgs potential.
Therefore, the mass degeneracy among the septet-like scalar bosons would be resolved. 
These mass splitting may give sizable contributions to the electroweak precision parameters 
as in the two Higgs doublet models~\cite{Ref:KOTT}. 
The potential parameters are further bounded by perturbative unitarity and 
the vacuum stability. 
If the singly charged Higgs boson is light, it may be constrained by flavor data. 
The Higgs sector with the Higgs septet potentially has rich phenomena. 
It would be worth investigating furthermore~\cite{Ref:HT}. 

\acknowledgments
 We thank Masaharu Tanabashi  for valuable comments.
This work is supported by Grant-in-Aid for Scientific research from 
the Ministry of Education, Science, Sports, and Culture (MEXT), Japan, 
No. 20244037, No. 20540252, No. 22244021 and No. 23104011, 
and also 
by World Premier International Research Center Initiative (WPI Initiative), MEXT, Japan.

\appendix
\section{One loop integrals}
The loop functions used in the evaluation of the oblique parameters are listed below,
\begin{align}
&F^{xy} =
\frac{m_x^2+m_y^2}2 -\frac{m_x^2m_y^2}{m_x^2-m_y^2}\ln\frac{m_x^2}{m_y^2}, \\
&G^{xV} = F^{xV} + 4\, m_V^2 
\Bigl( -1 +\frac{m_x^2 \ln m_x^2 -m_V^2 \ln m_V^2}{m_x^2-m_V^2}  \Bigr),\\
&\Delta G^x = G^{xW} -G^{xZ}, \\
&{F^{xy}}' = 
-\frac13 \Bigl( +\frac43
-\frac{m_x^2 \ln m_x^2 -m_y^2 \ln m_y^2}{m_x^2-m_y^2}
-\frac{m_x^2+m_y^2}{(m_x^2-m_y^2)^2}\, F^{xy} \Bigr)
, \\
&{G^{xV}}' = {F^{xV}}' + 4\, m_V^2 
\Bigl( -\frac1{(m_x^2-m_V^2)^2}\, F^{xV} \Bigr).
\end{align}



\begin{thebibliography}{99}

\bibitem{Ref:atlas}
  G.~Aad {\it et al.}  [ATLAS Collaboration],
  Phys.\ Lett.\ B {\bf 716}, 1 (2012). 

\bibitem{Ref:cms}
  S.~Chatrchyan {\it et al.}  [CMS Collaboration],
  Phys.\ Lett.\ B {\bf 716}, 30 (2012).

\bibitem{Ref:atlas-combi}
 The ATLAS Collaboration, 
  Report No. ATLAS-CONF-2012-170. 

\bibitem{Ref:cms-combi}
 The CMS Collaboration, 
  Report No. CMS-PAS-HIG-12-045. 

\bibitem{Ref:Type-II}
  W.~Konetschny and W.~Kummer,
  Phys.\ Lett.\  B {\bf 70}, 433 (1977);
%
  M.~Magg and C.~Wetterich,
  Phys.\ Lett.\  B {\bf 94}, 61 (1980);
%
  T.~P.~Cheng and L.~F.~Li,
  Phys.\ Rev.\  D {\bf 22}, 2860 (1980);
%
  J.~Schechter and J.~W.~F.~Valle,
  Phys.\ Rev.\  D {\bf 22}, 2227 (1980).

\bibitem{Ref:rho}
  D.~A.~Ross and M.~J.~G.~Veltman,
  Nucl.\ Phys.\ B {\bf 95}, 135 (1975);
%
  M.~J.~G.~Veltman,
  Nucl.\ Phys.\ B {\bf 123}, 89 (1977).
  
\bibitem{Ref:pdg}
  J.~Beringer {\it et al.} [Particle Data Group Collaboration], 
  Phys.\ Rev.\ D {\bf 86}, 010001 (2012).
  
\bibitem{Ref:GM} 
  H.~Georgi and M.~Machacek,
  Nucl.\ Phys.\ B {\bf 262}, 463 (1985);
%
  M.~S.~Chanowitz and M.~Golden,
  Phys.\ Lett.\ B {\bf 165}, 105 (1985).

\bibitem{Ref:HT}
J.~Hisano, K.~Tsumura, in preparation. 


\bibitem{Ref:PeskinTakeuchi}
  M.~E.~Peskin, T.~Takeuchi,
  Phys.\ Rev.\ Lett.\  {\bf 65}, 964-967 (1990); 
%
  Phys.\ Rev.\  {\bf D46}, 381-409 (1992).

\bibitem{Ref:HHG}
  J.~F.~Gunion, H.~E.~Haber, G.~Kane and S.~Dawson,
  The Higgs Hunter's Guide
  (Frontiers in Physics series, Addison-Wesley, Reading, MA, 1990).
  
\bibitem{Ref:XSWG} 
{\tt https://twiki.cern.ch/twiki/bin/view/LHCPhysics/CrossSections}

\bibitem{Ref:atlas-Dec4l}
 The ATLAS Collaboration, 
  Report No. ATLAS-CONF-2012-169. 

\bibitem{Ref:WZH-lhc}
  E.~Asakawa, S.~Kanemura and J.~Kanzaki,
  Phys.\ Rev.\ D {\bf 75}, 075022 (2007).

\bibitem{Ref:WZH-ilc}
  S.~Kanemura, K.~Yagyu and K.~Yanase,
  Phys.\ Rev.\ D {\bf 83}, 075018 (2011).

\bibitem{Ref:CNT}
  P.~Fileviez Perez, T.~Han, G.~-y.~Huang, T.~Li and K.~Wang,
  Phys.\ Rev.\ D {\bf 78}, 015018 (2008); 
%
  C.~-W.~Chiang, T.~Nomura and K.~Tsumura,
  Phys.\ Rev.\ D {\bf 85}, 095023 (2012)

\bibitem{Ref:KOTT}
  S.~Kanemura, Y.~Okada, H.~Taniguchi and K.~Tsumura,
  Phys.\ Lett.\ B {\bf 704}, 303 (2011).

\end{thebibliography}
\end{document}